\magnification\magstephalf
\baselineskip 14pt
\parskip 2pt

\def\pfbox
  {\hbox{\hskip 3pt
\vbox{\hrule\hbox to 7pt{\vrule height 7pt\hfill\vrule}
  \hrule}}\hskip3pt}

\def\bib{\par\noindent\hangindent 20pt}

\centerline{\bf Textbook Examples of Recursion}
\centerline{by Donald E. Knuth}

\medskip
{\narrower\smallskip\noindent
{\it Abstract}. We discuss properties of recursive schemas related
to McCarthy's ``91~function'' and to Takeuchi's triple recursion.
Several theorems are proposed as interesting candidates for machine
verification, and some intriguing open questions are raised.
\smallskip}
        
\bigskip
John McCarthy and Ikuo Takeuchi introduced interesting recurrence
equations as they were exploring the properties of recursive programs.
McCarthy's function~[7]
$$f(x)=\; {\bf if}\; x>100\; {\bf then}\; x-10\; {\bf else}\;
f\bigl(f(x+11)\bigr)$$
has become known as the ``91 function,'' since it turns out that
$f(x)=91$ for all $x\leq 101$. Takeuchi's function~[13] is a triple
recursion
$$t(x,y,z)= {\bf if}\;x\leq y\;
{\bf then}\; y\;
{\bf else}\;
t\bigl(t(x-1,y,z),\,t(y-1,z,x),\,t(z-1,x,y)\bigr)\,,$$
which has proved useful for benchmark testing of Lisp systems because
the recursion terminates only after the definition has been expanded
a~large number of times (assuming that previously computed values are
not remembered). Neither of these functions is of practical 
importance, because no  reasonable programmer would ever
want to carry out such recursive computations on a realistic
problem. Yet both functions are quite instructive because they
illustrate important problems and techniques that arise when we
consider
the task of verifying computer programs formally. Therefore they
make excellent examples for textbooks that discuss recursion.

The purpose of this paper is to obtain new information about
$f(x)$ and $t(x,y,z)$ and about several closely related functions.
Several of the theorems proved below should provide good test material
for automated verification systems. A~few open problems are stated,
illustrating the fact that extremely simple recursions can lead
to quite difficult questions.

\medskip
\noindent
{\bf 1. The 91 function.}\enspace
It is appropriate to begin by studying the 91 function, because
1991 is the year of John McCarthy's 64th birthday (and because a
computer scientist's most significant birthday is the 64th).
McCarthy originally wrote down the definition of
$f(x)$, as shown above, because he
wanted to study a simple recursion whose properties could not be
deduced by ordinary mathematical induction. After studying the
definition, he was pleasantly surprised to discover that it had the
totally unexpected ``91~property.'' 

The 91 function certainly belongs to the set of significant textbook
examples, because it is mentioned on at least 14~pages of Zohar
Manna's well known text, {\sl Mathematical Theory of
Computation\/}~[6].
The first published discussions of the function appeared in 1970
[7,~9], after it had been investigated extensively at Stanford's
AI~laboratory during 1968~[8].

Instead of using McCarthy's original definition, let's change the
specifications a bit and consider the following function (see~[6,
Problem~5--8]):
$$f(x)={\bf if}\;x>100\;{\bf then}\;x-10\;{\bf
else}\;f^{\mkern1mu 91}(x+901)$$
where $f^{\mkern1mu 91}(y)$ stands for 
$f\bigl(f\bigl(\,\cdots\,\bigl(f(y)\bigr)\,\cdots\,\bigr)\bigr)$, the
91-times-repeated application of~$f$.
According to this new definition, we have
$$f(91)=f^{\mkern1mu 91}(992)=f^{\mkern1mu 90}(982)=\cdots =f^2(102)=f(92)\,.$$
And a similar derivation shows that if $90\leq x\leq 100$ we have
$$f(x)=f^{\mkern1mu 91}(x+901)=\cdots =f^2(x+11)=f(x+1)\,;$$
hence
$$f(90)=f(91)=\cdots =f(100)=f(101)\,.$$
And $f(101)=91$, so we have proved in particular that
$$f(91)=91\,.$$

Now let's evaluate $f(x)$ when $x$ is extremely small, say $x=-10^6$.
We have
$$\eqalign{f(-1000000)=f^{\mkern1mu 91}(-999099)=f^{181}(-998198)=\cdots
&=f^{\mkern1mu 99811}(-791)\cr
&=f^{\mkern1mu 99901}(110)\cr
&=f^{\mkern1mu 99900}(100)\cr}$$
and we know that $f(100)=91$; hence
$$f^{\mkern1mu 99900}(100)=f^{\mkern1mu 99899}(91)
=f^{\mkern1mu 99898}(91)=\cdots =f(91)=91\,.$$
In general, if $x$ is any integer $\leq 100$, let $m$ be the smallest
integer such that $x+901m>100$, and let $n$ be the smallest integer
such
that $x+901m-10n\leq 100$. Then $m\geq 1$, $n\leq 91$, and
$$f(x)=f^{1+90m}(x+901m)=f^{1+90m-n}(x+901m-10n)=f^{\mkern1mu 90m-n}(91)\,,$$
where the last step follows since $91\leq x+901m-10n\leq 100$. We
conclude that $f(x)=91$.
(The final step is omitted if $m=1$ and $n=91$;  that case occurs
if and only if $x=100$.)

How many iterations are needed to compute $f(x)$  by this definition,
if we continue to apply the recurrence even when evaluating $f(x)$ for
values of~$x$ that have already been considered? Let $F(x)$ count the
number of times that
the test `{\bf if} $x>100$' is performed; then we have
$$\eqalign{F(x)={\bf if}\; x>100\;{\bf then}\;1\;{\bf else}\;1
+F&(x+901)+F\bigl(f(x+901)\bigr)+\cr
\noalign{\smallskip}
&\;\null +F\bigl(f^2(x+901)\bigr)+\cdots
+F\bigl(f^{\mkern1mu 90}(x+901)\bigr)\,.\cr}$$
(This is a special case of the general notion of a derived function,
which
is always jointly recursive with the function from which it has been
derived; see McCarthy and Talcott~[11].) A~bit of
experimentation reveals that $F(x)$ also reduces to a simple function:

\proclaim Lemma 1. $F(x)=$ {\bf if} $x>100$ {\bf then} $1$ {\bf else}
$9192-91x$.

\noindent 
{\it Proof}. \enspace If $x<100$ we have
$$F(x)-F(x+1)=\sum_{k=0}^{\mkern1mu 90}\bigl(F\bigl(f^k(x+901)\bigr)
-F\bigl(f^k(x+902)\bigr)\bigr)\,.$$
Now if $x+901\leq 100$, the sum reduces to $F(x+901)-F(x+902)$,
because
 the terms for $k>0$ are
$F\bigl(f^k(x+901)\bigr)-F\bigl(f^k(x+902)\bigr)
=F(91)-F(91)=0$.
In this case we let $x'=x+901$. On the other hand if
$x+901>100$, let $n$ be minimal such that
$x+901-10n\leq 100$. Then $1\leq n\leq 90$, and
$F\bigl(f^k(x+901)\bigr)-F\bigl(f^k(x+902)\bigr)=
F(91)-F(91)=0$ 
 for all $k>n$.  We also have $F\bigl(f^n(x+901)\bigr)
-F\bigl(f^n(x+902)\bigr)=F(x+901-10n)-F(x+902-10n)$; and
$F\bigl(f^k(x+902)\bigr)-F\bigl(f^k(x+901)\bigr)=1-1=0$ for all
$k<n$. In this case we let $x'=x+901-10n$.
In both cases we have found an~$x'$ such that
$$F(x+1)-F(x)=F(x'+1)-F(x')\,,\qquad x<x'\leq 100\,.$$
The proof is therefore complete by induction on $101-x$ if we
simply verify that $F(100)-F(101)=91$.\quad\pfbox

\medskip
The 91 function suggests that we consider the more general
recursive scheme
$$f(x)={\bf if}\;x>a\;{\bf then}\;x-b\;{\bf else}\;f^c(x+d)\,,$$
where $a$ is an arbitrary real number, $b$ and $d$ are positive reals,
and~$c$ is a positive integer.

\proclaim Theorem 1.
The generalized 91 recursion with parameters $(a,b,c,d)$ defines a
total function on the integers if and only if $\,(c-1)\,b<d$. In such a
case the values of $f(x)$ also obey the much simpler recurrence
$$f(x)={\bf if}\;x>a\;{\bf then}\;x-b\;{\bf else}\;f\bigl(x+d-(c-1)\,b\bigr)
\,.$$

\noindent
{\it Proof}.\enspace
It is not difficult to show that any function satisfying the
generalized
91 recursion for $c>1$ must also satisfy
$$f(x)={\bf if}\;x>a\;{\bf then}\;x-b\;{\bf else}\;f^{c-1}(x+d-b)\,.$$
For if $x\leq a$, let $n$ be minimal such that $x+nd>a$. Then
$$\displaylines{f^c(x+d)=f^{c+(n-1)(c-1)}(x+nd)=f^{n(c-1)}(x+nd-b)\,;\cr
\noalign{\smallskip}
f^{c-1}(x+d-b)=f^{c-1+(n-1)(c-1)}\bigl(x+d-b+(n-1)\,d\bigr)\,;\cr}$$
hence $f^c(x+d)=f^{c-1}(x+d-b)$, as desired.

To complete the proof, we 
use induction on~$c$; and we also
need to characterize the parameter settings
that cause the given recursive definition to terminate for all~$x$.

If $(c-1)\,b\geq d$, the expansion of $f(x)$  will not terminate when
$a-b<x\leq a$. For if $n$ is minimum such that $x+d-nb\leq a$, we have
$n\leq c-1$, and
$$f(x)=f^c(x+d)=\cdots =f^{c-n}(x+d-nb)\,.$$
Now $c-n>0$, and $a-b<x+d-nb\leq a$, so this will go on and~on.

On the other hand, we can show that no looping will occur if
$(c-1)\,b<d$. Suppose first that $x>a-b$. If $x>a$, obviously
$f(x)=x-b$. Otherwise we have $x+d>x+(c-1)\,b>a+(c-2)\,b$, hence
$$f(x)=f^c(x+d)=\cdots
=f^2\bigl(x+d-(c-2)\,b\bigr)=f\bigl(x+d-(c-1)\,b\bigr)\,.$$
Let $\Delta=d-(c-1)\,b$, and let $m$ be minimum such that
$x+m\,\Delta>a$; then
$$f(x)=f(x+\Delta)=\cdots =f(x+m\,\Delta)=x+m\,\Delta-b\,.$$
Thus, the expansion of $f(x)$ terminates with a value $>a-b$ whenever
$x>a-b$. 

Finally, if $x\leq a-b$ and if $m$ is minimal such that $x+md>a-b$,
the expansion of
$$f(x)=f^{1+m(c-1)}(x+md)$$
terminates, because we can peel off the $f$'s one by one.\quad\pfbox

When the generalized 91 function is total, we can express it in
``closed form''~as
$$\eqalign{f(x)=\;&{\bf if}\;x>a\;{\bf then}\;x-b\cr
&{\bf
else}\;a+d-cb-\bigl((a-x)\bmod\bigl(d-(c-1)\,b\bigr)\bigr)\,.\cr}$$
The special case $c=2$ of Theorem 1 was first proved by Manna and
Pnueli~[8]. 

\medskip\noindent
{\bf Open Problem 1.}\enspace
Prove Theorem 1 by computer.\quad\pfbox

\bigskip\noindent
{\bf 2. The Takeuchi function.}\enspace
Now we turn to the more complex recurrence
$$t(x,y,z)={\bf if}\;x\leq y\;{\bf then}\;y\;{\bf else}\;
t\bigl(t(x-1,y,z),\,t(y-1,z,x),\,t(z-1,x,y)\bigr)\,.$$
John McCarthy observed in unpublished notes~[10]
that this function can be
described more simply as
$$t(x,y,z)={\bf if}\;x\leq y\;{\bf then}\;y\;{\bf else}\;{\bf if}\;
y\leq z\;{\bf then}\; z\;{\bf else}\;x\,.$$
If we assume termination, the latter function satisfies Takeuchi's 
recurrence, so it must be identical with the former function.

John had just returned from a conference in Kyoto, and his notes~[10]
began with a brief comment about the history of this function and its
motivation:

{\narrower\smallskip\noindent
Ikuo Takeuchi (1978) of the Electrical Communication Laboratory of
Nippon Telephone and Telegraph Co.\ (Japan's Bell Labs) devised a
recursive function program for comparing the speeds of LISP systems.
It can be made to run a long time without generating large numbers or
using much stack.
\smallskip}

\noindent
(Incidentally, I believe~[10] was John's first experiment with the use
of \TeX, a~computer typesetting system that I~was developing while
sitting in the office next to his. Without his generous provision of
computing and printing facilities, \TeX\ would never have existed.)

At about the same time, John coerced the FOL proof-checking system to
construct a 50-step proof that $A(x,y,z)$ has the simple form stated
above~[10]. This experiment suggested several improvements to FOL.

If we fully expand the definition of $t(x,y,z)$ whenever $x>y$, the
 proof of termination seems to be nontrivial, because there is no
obvious way to impose an order on the set of all arguments $(x,y,z)$ 
in such a way
that no infinitely long dependency chains exist. We shall prove
termination
as a byproduct of a more general investigation of the total running
time
needed to evaluate $t(x,y,z)$ by repeated application of the
definition.

Let $T(x,y,z)$ be the number of times the {\bf else} clause is invoked
when $t(x,y,z)$ is evaluated by ordinary Lisp recursion. Then
$$\eqalign{T(x,y,z)=\;&{\bf if}\;x\leq y\;{\bf then}\;0\cr
&{\bf else}\;1+T(x-1,y,z)+T(y-1,z,x)+T(z-1,x,y)\cr
&\phantom{{\bf else}\;1}\null +T\bigl(t(x-1,y,z),\,t(y-1,z,x),
\, t(z-1,x,y)\bigr)\,.\cr}$$
The total number of expansions of the definition will then be
$1+4T(x,y,z)$, because the latter function ${\cal T}(x,y,z)$ satisfies
the recurrence 
$$\eqalign{{\cal T}(x,y,z)=\;&{\bf if}\;x\leq y\;{\bf
then}\;1\cr 
&{\bf else}\;1+{\cal T}(x-1,y,z)+{\cal T}(y-1,z,x)+{\cal
T}(z-1,x,y)\cr
&\phantom{{\bf else}\;1}\null +{\cal
T}\bigl(t(x-1,y,z),\,t(y-1,z,x),\,t(z-1,x,y)\bigr)\,.\cr}$$ 

Before we analyze $T(x,y,z)$ it will be helpful to consider a similar 
but simpler function
$$\eqalign{V(x,y,z)=\;&{\bf if}\;x\leq y\;{\bf then}\;0\cr
&{\bf else}\;1+V(x-1,y,z)+V(y-1,z,x)+V(z-1,x,y)\,.\cr}$$
The function $V(x,y,z)$ can be understood as follows.
Construct a ternary tree by starting with a simple leaf containing
 the triple $[x,y,z]$ 
and repeatedly applying the following operation: If any leaf $[x,y,z]$
of the tree-so-far has $x>y$, attach the nodes
$$[x-1,y,z]\,,\quad [y-1,z,x]\,,\quad [z-1,x,y]$$
immediately below it. Then $V(x,y,z)$ will be the number of nonleaf
nodes in the final tree. (This function $V(x,y,z)$ has been studied
by Ilan Vardi~[14]; some of his analysis is reproduced here.)

[Note: I have an example in the MS.,  but I'll skip it unless I get goahead
from Vlad later.]

The evaluation of $V(x,y,z)$ is trivial when $x\leq y$, and it's also
fairly simple when $x>y$ and
$x\geq z\geq y$: In that case we have
$$V(x,y,z)=1+V(x-1,y,z)=1+x-z+V(z-1,y,z)\,.$$

A further simplification arises when we realize that the values of $V(x,y,z)$
are invariant if we translate all the parameters by any integer
amount:
$$V(x+1,y+1,z+1)=V(x,y,z)\,.$$
Therefore we can shorten our notation and our discussion by assuming
that
$\min(x,y,z)=0$. 

Suppose the ternary tree has $[x,y,z]$ at the root, where
$\min(x,y,z)=0$ and either $x>y>z$ or $z>x>y$. Then all of its
non-leaf nodes are of two kinds,
$$\matrix{A(a,b)=[a,b,0]\cr
\noalign{\smallskip}
B(a,b)=[b,0,a]\cr}\qquad{\rm where}\;a>b>0\,.$$
Below $A(a,b)$ are the three nodes
$$[a-1,b,0]\,,\quad [b-1,0,a]\,,\quad [-1,a,b]$$
where $[a-1,b,0]$ is a leaf if $a=b+1$, otherwise it is $A(a-1,b)$;
similarly $[b-1,0,a]$ is a leaf if $b=1$, otherwise it is $B(a,b-1)$;
and $[-1,a,b]$ is always a leaf. Below $B(a,b)$ are the same three 
nodes; they appear in a different order, but that does not matter.

It follows that $V(x,y,0)=V(y,0,x)$, for all  $x>y>0$,
and that $V(x,y,0)$
 has a simple
combinatorial interpretation: It is the number of lattice paths that
start at $(x,y)$ and stay within the set $\{\,(a,b)\mid a>b>0\,\}$.
(A~lattice path is a path in which each step decreases exactly one
of the coordinates by unity.)  We will say that such a lattice path
is {\it confined}.

Confined lattice paths can be enumerated by using Andr\'e's well-known
reflection principle (see, for example, [4, exerise 2.2.1--4]).
Given $x>y\geq y'>0$, the number of confined paths from $(x,y)$
to a point $(x',y')$ for some~$x'$ is equal to the number of all
possible lattice paths from $(x,y)$ to $(y',y'-1)$ minus the number of
such paths that touch a diagonal point. Paths of the latter type are
in
one-to-one correspondence with lattice paths from $(x,y)$ to
$(y'-1,y')$; the correspondence is obtained by interchanging $x$~moves
with $y$~moves after the diagonal is first encountered.
Hence the number of
confined paths from $(x,y)$ to points of the form $(x',y')$, given
$x$, $y$, and~$y'$, is
$$\pmatrix{x-y'+y-(y'-1)\cr x-y'\cr}-\pmatrix{x-(y'-1)+y-y'\cr
y-y'\cr}\,;$$
and the total number of confined paths starting from $(x,y)$ is
$$\eqalign{%
&\sum_{y'=1}^y\,\left(\pmatrix{x+y+1-2y'\cr x-y'}-
\pmatrix{x+y+1-2y'\cr y-y'}\right)\cr
\noalign{\medskip}
&\qquad =
\sum_{k=1}^y\,\left(\pmatrix{x-y-1+2k\cr k}-
\pmatrix{x-y-1+2k\cr k-1}\right)\,.\cr}$$
This is the quantity $V(x,y,0)$. Notice that we have
$$V(n+1,n,0)=\sum_{k=1}^n\,\left(\pmatrix{2k\cr k\cr}-
\pmatrix{2k\cr k-1\cr}\right)=\sum_{k=1}^n\,{2k\choose k}\,
{1\over k+1}=\sum_{k=1}^n\,C_k\,,$$
the sum of the first $n$ Catalan numbers.

Returning to the evaluation of $T(x,y,z)$, we must add to $V(x,y,z)$
the values $T\bigl(t(a-1,b,0)$,
$t(b-1,0,a),a\bigr)$ at every node of
type $A(a,b)$, and the values $T\bigl(t(b-1,0,a),a,t(a-1,b,0)\bigr)$
at every node of type $B(a,b)$. Fortunately these additional amounts
are mostly zero. Our knowledge about the values of $t(x,y,z)$ allows
us to conclude that, when $a>b>0$, we have
$$t(a-1,b,0)=a-1\,;\qquad t(b-1,0,a)=\cases{0\,,&$b=1$;\cr
a\,,&$b>1$.\cr}$$
Therefore $T\bigl(t(b-1,0,a),\,a,\,t(a-1,b,0)\bigr)=0$; and
$T\bigl(t(a-1,b,0),\,t(b-1,0,a),\,a\bigr)=0$ except when
$b=1$. Only the nodes of type $A(a,1,0)$ acquire additional values;
and at such nodes we add $T(a-1,0,a)$.

The number of non-root nodes of type $A(a',1)$ in a tree whose root is
of
type $A(a,b)$ or $B(a,b)$ is the number of nodes of type $A(a'+1,1)$
or $B(a'+1,1)$. And the number of such nodes is the number of 
confined lattice paths from $(a,b)$ to $(a'+1,1)$, which is
$${a-(a'+1)+b-1\choose b-1}-{a-1+b-(a'+1)\choose a-1}$$
by Andr\'e's reflection principle. Therefore
$$T(b,0,a)=V(a,b,0)+\sum_{a'=2}^{a-1}\,\left({a+b-a'-2\choose b-1}
-{a+b-a'-2\choose a-1}\right)\,T(a'-1,0,a')\,.$$
The same formula holds for $T(a,b,0)$, except that we must add
$T(a-1,0,a)$ when $b=1$; a~root node of type $A(a,1)$ makes a
contribution, but a root node of type $B(a,1)$ does not.
Putting these facts together yields the following recurrence for the
numbers $T_n=T(n,0,n+1)$:
$$T_{n+1}=V(n+2,n+1,0)+\sum_{k=0}^{n-1}\,\left({n+k\choose n}-
{n+k\choose n+1}\right)\,T_{n-k}\,.$$
Let $V_n=V(n+1,n,0)$; the first few values of these sequences are as
follows:
$$\vcenter{\halign{\hfil$#\;$%
&$\hfil#$\enspace&$\hfil#$\enspace&$\hfil#$\enspace&$\hfil#$\enspace%
&$\hfil#$\enspace&$\hfil#$\enspace&$\hfil#$\enspace&$\hfil#$\enspace%
&$\hfil#$\cr
n=&1&2&3&4&5&6&7&8&9\cr
\noalign{\smallskip}
V_n=&1&3&8&22&64&196&625&2055&6917\cr
\noalign{\smallskip}
T_n=&1&4&14&53&223&1034&5221&28437&165859\cr}}$$
It is not difficult to deduce that the numbers $T_n$ grow very
rapidly,
in fact faster than~$A^n$ for any constant~$A$:

\proclaim Lemma 2. If $\epsilon >0$, we have
$$T_n>n^{(1-\epsilon)n}$$
for all sufficiently large $n$.

\noindent
{\it Proof}. \enspace Choose $k$ large enough so that $k/(k+1)>1-\epsilon$. 
Looking only at the $k$th~term of the recurrence for~$T_n$
tells us that, for all $n>k$, we have
$$T_{n+1}>\left({n+k\choose n}-{n+k\choose n+1}\right)\,T_{n-k}
={(n+k)(n+k-1)\ldots (n+2)(n+1-k)\over k!}\,T_{n-k}\,.$$
Thus
$$\ln T_{n+1}>\ln(n+k)+\cdots +\ln(n+2)+\ln(n+1-k)-\ln k!+\ln T_{n-k}\,.$$
Iterating this relation yields
$$\ln T_n>{k\over k+1}\,n\ln n+O(n)$$
and the result follows. $\bigl($A~similar but weaker result was obtained by
Ilan Vardi~[14], who used the fact that $T_{n+1}>n\,T_{n-1}$ to prove that
$\ln T_n>{1\over 2}n\ln n+O(n)$.$\bigr)$\quad\pfbox

\medskip
In fact, we can prove a stronger lower bound by observing that for all
$n\geq 1$, we have $T_n\geq b_n$, where $b_n$ is the Bell number defined by
$$b_{n+1}=1+\sum_{k=0}^{n-1}\,{n\choose k}\,b_{n-k}\,.$$
Each term in the recurrence for $T_n$ is greater than or equal to the
corresponding term in the recurrence for~$b_n$, by induction. It is
known
[1,~(6.2.7)] that
$$b_n>e^{n\ln n-n\ln\ln n-n}$$
for all sufficiently large  $n$. Thus $T_n$ grows faster than
$(n/e\ln n)^n$.

On the other hand, we can prove the upper bound
$$T_n<3n!\;.$$
This is clear for $n\leq 3$. Let $t_n=T_n/n!$, and rewrite the
recurrence
for $T_n$ as follows:
$$\eqalign{%
t_{n+1}={V_{n+1}\over (n+1)!}+{1\over n+1}\,&\left(t_n+t_{n-1}+
{n+2\over 2n}\,t_{n-2}\right.\cr
\noalign{\medskip}
&\qquad\null
+\left.{n+3\over 3(n-1)}\,{n+2\over 2n}\,t_{n-3}
+{n+4\over 4(n-2)}\,{n+3\over 3(n-1)}\,{n+2\over 2n}\,t_{n-4}
+\cdots\;\right)\,.\cr}$$
The coefficients inside the parentheses are all $\leq 1$, so we have
$$t_{n+1}\leq {V_{n+1}\over (n+1)!}+{1\over n+1}\,(t_n+t_{n-1}+
\cdots +t_1)<{3n\over n+1}+{V_{n+1}\over (n+1)!}$$
by induction. And it is easy to verify that $V_{n+1}<3n!$ for 
$n\geq 3$, because $V_{n+1}\leq 4^n$. (The Catalan numbers~$C_n$
satisfy $C_{n+1}<4C_n$, hence $V_{n+1}<C_1+4V_n$ and we have
$V_{n+1}\leq 4V_n$.) We have proved

\proclaim Theorem 2. When the Takeuchi recursion $t(x,y,z)$ is used
in a memoryless manner to evaluate $t(n,0,n+1)$, the definition is
expanded $1+4T(n,0,n+1)$ times, where
$$e^{n\ln n-n\ln\ln n-n}<T(n,0,n+1)<e^{n\ln n-n+\ln n}$$
for all sufficiently large $n$.\quad\pfbox

\medskip
\noindent
Only $O(n^2)$ evaluations are needed when previously computed results
are remembered; thus memory is especially helpful here.

A more precise asymptotic formula for $V_n$ can be obtained from the
well-known generating function for Catalan numbers [3,
page~203],
$$\displaylines{V(z)=\sum_{n\geq 1}\,V_nz^n={1\over 1-z}\,
\sum_{n\geq 1}\,C_nz^n={C(z)-1\over 1-z}\,,\cr
\noalign{\smallskip}
C(z)={1-\sqrt{1-4z}\,\over 2z}\,.\cr}$$
Darboux's lemma (see~[5]) now shows that
$${V_n\over 4^n}=[z^n]\,V\left({1\over 4}\,z\right)=-{8\over 3}\,
{n-3/2\choose n}+O(n^{-5/2})={4n^{-3/2}\over 3\sqrt{\pi}}+
O(n^{-5/2})\,.$$

The generating function of the numbers $T_n$ satisfies a remarkable
functional equation: We have
$$\eqalign{T(z)&=\sum_n\,T_nz^n\cr
\noalign{\medskip}
&=\sum_n\,V_{n+1}z^{n+1}=\sum_{n,k}\,{n+k\choose k}\,T_{n-k}z^{n+1}
-\sum_{n,k}\,{n+k\choose k-1}\,T_{n-k}z^{n+1}\cr
\noalign{\medskip}
&={C(z)-1\over 1-z}+\sum_{n,k}\,{n+2k\choose k}\,T_nz^{n+k+1}
-\sum_{n,k}\,{n+2k\choose k-1}\,T_nz^{n+k+1}\cr
\noalign{\medskip}
&={C(z)-1\over 1-z}+\sum_n\,T_nz^{n+1}\,{C(z)^n\over\sqrt{1-4z}\,}
-\sum_n\,T_nz^{n+2}\,{C(z)^{n+2}\over\sqrt{1-4z}\,}\cr
\noalign{\medskip}
&={C(z)-1\over 1-z}+{z\bigl(2-C(z)\bigr)\over \sqrt{1-4z}}\,
T\bigl(zC(z)\bigr)\,,\cr}$$
because $zC(z)^2=C(z)-1$ and $\sum_k{n+2k\choose k}\,z^k
=C(z)^k/\sqrt{1-4z}$.

\bigskip\noindent
{\bf Open Problem 2.}\enspace
Obtain further information about the asymptotic properties of the
coefficients
$T_1,T_2,\ldots\;$.\quad\pfbox

\medskip
The evaluation of $t(x,y,z)$ turns out to be much, much faster if we
apply the technique of lazy evaluation or ``call by need'' when
expanding the definition. (See Vuillemin [15], [16].) Indeed, we
can ignore the third argument $t(z-1,x,y)$ in the recursion, unless we
have discovered that $t(x-1,y,z)>t(y-1,z,x)$; so the number of times
the {\bf else} clause needs to be expanded satisfies the recursion
$$\eqalign{K(x,y,z)=\;&{\bf if}\; x\ge	 y\;{\bf then}\;0\cr
&{\bf else}\;\bigl(1+K(x-1,y,z)+K(y-1,z,x)\cr
&\phantom{{\bf else}\;\bigl(}\null+\;{\bf if}\;t(x-1,y,z)\leq
t(y-1,z,x)\;{\bf then}\;0\cr
&\phantom{{\bf else}\;\bigl(\null+\,\;\;}{\bf else}\;
K(z-1,x,y)+K\bigl(t(x-1,y,z),\,t(y-1,z,x),\,t(z-1,x,y)\bigr)\bigr)\,.\cr}$$
And this recursion turns out to be quite simple. First, if $x>y\leq
z$, we have
$$K(x,y,z)=1+K(x-1,y,z)=x-y$$
because $t(x-1,y,z)\leq z$ and $t(y-1,z,x)=z$. Second, if $x>y>z+1$,
we have
$$\eqalign{K(x,y,z)&=1+K(x-1,y,z)+K(y-1,z,x)\cr
&=1+K(x-1,y,z)+y-1-z=(x-y)(y-z)\,,\cr}$$
because $t(x-1,y,z)=x-1$ and $t(y-1,z,x)=x$. Finally, if $x>y=z+1$, we
have $t(x-1,y,z)=x-1$, $t(y-1,z,x)=z$, and $t(z-1,x,y)=x$; hence
$$K(x,y,z)=1+K(x-1,y,z)+x-1-z=(x-y)(x-y+3)/2\,.$$

Incidentally, when $x>y>z+1$, the expansions of {\bf else} clauses
occur at the arguments $(\xi,y,z)$ and $(\eta,z,\xi)$ for $x\geq\xi>y$
and $y>\eta>z$; when $x>y=z+1$, they occur at $(\xi,y,z)$ and
$(\eta,z,\xi)$ for $x\geq\xi>\eta\geq y$. Since these arguments are
distinct, no additional savings over call-by-need would be obtained by
remembering previously computed values, unless $t(x,y,z)$ is being
evaluated at more than one point $(x,y,z)$. The fact that the
necessary arguments are limited underlies the simple mechanical proof
of termination found by Moore~[12].

\bigskip\noindent
{\bf 3. False Takeuchi functions.}\enspace
Vardi [14] has considered a general recursion scheme of the form
$$v_h(x,y,z)={\bf if}\; x\leq y \;{\bf then}\; h(x,y,z)\;
{\bf else}\;
v_h\bigl(v_h(x-1,y,z),\,v_h(y-1,z,x),\,v_h(z-1,x,y)\bigr)\,.$$
If we set $h(x,y,z)=0$, the function $v_h(x,y,z)$ will of course be
identically zero; we will deduce that its value is zero
 after expanding the
definition exactly $1+4V(x,y,z)$ times, where $V(x,y,z)$ is the
function considered in the previous section. This is clearly the
minimum
number of expansions necessary over all possible auxiliary functions
$h(x,y,z)$, if we do not or cannot use call-by-need.

Richard Gabriel used a Takeuchi-like function in extensive benchmark tests of
 Lisp compilers, but his function was slightly different from
Takeuchi's original:
$$g(x,y,z)={\bf if}\;x\leq y\;{\bf then}\;z\;{\bf else}\;
g\bigl(g(x-1,y,z),\,g(y-1,z,x),\,g(z-1,x,y)\bigr)\,.$$
(Notice that in this case call-by-need is inapplicable.) Gabriel
explains the discrepancy as follows [2,~pp.\ 10--11]:

{\narrower\smallskip\noindent
When the Computer Science Department at Stanford University obtained
the first two or three Xerox Dolphins, John McCarthy asked me to do a
simple benchmark test with him. We sat down, and he tried to remember
the Takeuchi function, which had had wide circulation. Because it was
simple and because there were many results for it in the literature,
he felt that it would be a good initial test. Of course, John
misremembered the function. But we did not realize it until I~had
gathered a great many numbers for~it.
\smallskip}

\noindent
Indeed, Gabriel's book [2] gives detailed timings for the
computation of $g(18,12,6)=7$ on 132~different configurations, and he
lists four additional variants of~$g$ that provide further types of
benchmark tests.

The seemingly trivial change from $t(x,y,z)$ to $g(x,y,z)$ actually
makes $g(x,y,z)$ substantially easier to compute, if $t(x,y,z)$ is not
evaluated with memory of previous results or with call-by-need. Vardi~[14]
 has shown that the corresponding running time $G(n,0,n+1)$ is
asymptotically less than $(3+\sqrt{8}\,)^n$, although the exact
order of growth is not known.
Vardi has also observed that Gabriel's recursion defines the following
curious pattern of values:

\smallskip
\halign{\qquad\qquad#\hfil&#\hfil\cr
$g(x,y,z)=\;$&{\bf if} $x\leq y$ {\bf then} $z$\cr
&{\bf else}$\;${\bf if} $y\geq z$ {\bf then}\cr
&\qquad\qquad$\!${\bf if} $y=z$ or $(x-y)$ odd {\bf then}
$y$\cr
&\qquad\qquad$\!${\bf else} $z+1$\cr
&{\bf else if} $z\leq x+1$ and ($z\leq x$ or $x>y+1$) {\bf then}
$y$\cr
&\qquad$\,${\bf else if} $(z-x)$ even {\bf then} $x$\cr
&\qquad$\,${\bf else} $y+1\,$.\cr}

Here is another example of a generalized Takeuchi recurrence whose
solution  exhibits odd-even behavior:

\smallskip
\halign{\qquad\qquad$#\;$\hfil&#\hfil\cr
b(x,y,z)=\;&{\bf if} $x\leq y$ {\bf then if} $x=y=z$ {\bf then} 0
{\bf else} 1\cr
&{\bf else} $b\bigl(b(x-1,y,z),\,b(y-1,z,x),\,b(z-1,x,y)\bigr)\,.$\cr}

\smallskip\noindent
This time the output of the function is boolean---always either 0
or~1---although $x,y,z$ range over all integers. The computed values turn
out to~be

\smallskip
\halign{\qquad\qquad$#\;$\hfil&#\hfil\cr
b(x,y,z)=\;&{\bf if} $x\leq y$ {\bf then if} $x=y=z$ {\bf then} 0
{\bf else} 1\cr
&{\bf else if} $z>y+1$ {\bf then if} $(x-z)$ odd {\bf then} 0
{\bf else} 1\cr
&{\bf else if} $y=z$ {\bf then if} $(x-y)$ even {\bf then} 0
{\bf else} 1\cr
&{\bf else if} $(x-y)$ odd {\bf then} 0 {\bf else} 1.\cr}

\smallskip
The generalized recursion $v_h$ does not always define a total
function by repeated expansion.
For example, consider the auxiliary function
$$e(x,y,z)={\bf if}\; x\;{\rm odd}\;{\bf then}\;0\;{\bf else}\;1;$$
then we get
$$v_e(1,0,0)=v_e\bigl(v_e(0,0,0),\,v_e(-1,0,1),\,v_e(-1,1,0)\bigr)
=v_e(1,0,0)$$
and the recursion loops endlessly. There is a simple characterization
of the cases where $v_h$ is total in the boolean case:

\proclaim Lemma 4. Let $h(x,y,z)$ map arbitrary integers $x,y,z$
into 0 or~1. Then the recursive equation for $v_h(x,y,z)$ defines
a total function~$v_h$ except in the following three cases:

\halign{\qquad\hfil#\quad&{\sl #}\hfil\cr
(i)&$h(0,0,0)=1$ and $h(-1,0,1)=h(-1,1,0)=0$;\cr
\noalign{\smallskip}
(ii)&$h(0,0,1)=h(0,1,0)=1$ and $h(-1,1,1)=0$;\cr
\noalign{\smallskip}
(iii)&$h(0,0,0)=h(0,0,1)=h(-1,1,0)=1$ and $h(-1,0,1)=h(-1,1,1)=0$.\cr}

\bigskip\noindent
{\it Proof}.\enspace
If $h(0,0,0)=0$ or $h(-1,0,1)=1$, we have the well-defined result
$$v_h(1,0,0)=h\bigl(h(0,0,0),\,h(-1,0,1),\,h(-1,1,0)\bigr)\,;$$
otherwise we have
$$v_h(1,0,0)=v_h\bigl(1,0,h(-1,1,0)\bigr)\,,$$
which loops in case (i) but gives $v_h(1,0,0)=v_h(1,0,1)$ otherwise.
Similarly, we find that $h(0,0,1)=0$ or $h(-1,1,1)=1$ implies
$$v_h(1,0,1)=h\bigl(h(0,0,1),\,h(-1,1,1,\,h(0,1,0)\bigr)\,;$$
hence $v_h(1,0,1)$ is well defined whenever case (ii) does not hold,
except in the case when it leads to $v_h(1,0,1)=v_h(1,0,0)$.

If neither case (i) nor case (ii) holds, then $v_h(x,y,z)$ is well
defined for all boolean values $x,y,z$, except in case~(iii).
And when $v_h(x,y,z)$ is defined for all boolean $x,y,z$, we can
evaluate $v_h(x,y,z)$ for all $x,y,z$ in $O\bigl(V(x,y,z)\bigr)$
steps.\quad\pfbox

\smallskip
When the boolean function $v_h$ of Lemma 3 isn't total, we can always
complete it to a total function $v_h(x,y,z)$ that does satisfy
the recurrence. We simply assign arbitrary boolean values to
$v_h(1,0,0)$ and/or $v_h(1,0,1)$, whichever is undefined. For example,
there are four total functions $v_e(x,y,z)$ that satisfy the
recurrence
 arising from the auxiliary function $e(x,y,z)$ considered above:

\smallskip
\halign{\qquad\qquad$\hfil#\;$&#\hfil\cr
v_k^{00}(x,y,z)=\;&{\bf if} $x$ odd {\bf then} 0 {\bf else} 1;\cr
\noalign{\medskip}
v_e^{01}(x,y,z)=\;&{\bf if} $x\leq y$ {\bf then if} $x$ odd {\bf then} 0
{\bf else} 1\cr
&{\bf else if} $x$ even {\bf then} 1\cr
&{\bf else if} $y$ odd {\bf then} 0\cr
&{\bf else if} $z$ odd {\bf then} 1 {\bf else} 0;\cr
\noalign{\medskip}
v_e^{10}(x,y,z)=\;&{\bf if} $x\leq y$ {\bf then if} $x$ odd {\bf then}
0 {\bf else} 1\cr
&{\bf else if} $x$ odd {\bf then}\cr
&\qquad\qquad{\bf if} $y$ odd or $z$ odd {\bf then} 0 {\bf else} 1\cr
&{\bf else if} $y$ odd or $z$ odd {\bf then} 1\cr
&{\bf else if} $y\leq z\leq x$ {\bf then} 1 {\bf else} 0;\cr
\noalign{\medskip}
v_e^{11}(x,y,z)=\;&{\bf if} $x\leq y$ {\bf then if} $x$ odd {\bf then} 0
{\bf else} 1\cr
&{\bf else if} $x$ even {\bf then} 1\cr
&{\bf else if} $y$ odd {\bf then} 0\cr
&{\bf else if} $z\leq y$ {\bf then} 1\cr
&{\bf else if} $z$ odd {\bf then} 1 {\bf else} 0.\cr}

However, the recurrence $v_h(x,y,z)$ cannot be completed to a total
function for {\it arbitrary\/} auxiliary functions $h(x,y,z)$. Consider,
for example, the (admittedly contrived) mapping
$$h(x,y,z)=2xy-4x+y+z-1\,.$$
There is no total function $v_h$ because we would otherwise have
$$\eqalign{v_h(2,1,4)&=v_h\bigl(h(1,1,4),\;h(0,4,2),\;v_h\bigl(h(2,2,1),\,
h(1,1,3),\,h(0,3,2)\bigr)\bigr)\cr
\noalign{\smallskip}
&=v_h\bigl(2,5,v_h(2,1,4)\bigr)=16+v_h(2,1,4)\,.\cr}$$

An incompletable function can even be constructed when we restrict
ourselves to auxiliary functions that are limited by the condition
$$h(x,y,z)\leq \max(x,y,z)\,.$$
For example, suppose we define
$$\eqalign{h(x,y,z)=\;&{\bf if}\; (x,y,z)=(1,1,4)\;{\bf then}\;4\cr
&{\bf else\ if}\;(x,y,z)=(3,3,3)\;{\bf then}\;2\cr
&{\bf else\ if}\;(x,y)=(2,3)\;{\bf then}\;1\cr
&{\bf else\ if}\;\max(x,y,z)\geq 3\;{\bf then}\;3\cr
&{\bf else}\;\max(x,y,z)\,.\cr}$$
Then we have $v_h(x,y,3)=3$ whenever $3>x\geq y$. For if $x=y$,
clearly
$v_h(y,y,3)=h(y,y,3)=3$; otherwise
$$\eqalign{v_h(x,y,3)&=v_h\bigl(v_h(x-1,y,3),\,v_h(y-1,3,x),\,
v_h(2,x,y)\bigr)\cr
\noalign{\smallskip}
&=v_h(3,3,\leq 2)=3\,.\cr}$$
Therefore if $y<3$ we have
$$\eqalign{v_h(3,y,3)&=v_h\bigl(v_h(2,y,3),\,v_h(y-1,3,3),\,
v_h(2,3,y)\bigr)\cr
\noalign{\smallskip}
&=v_h(3,3,1)=3\,.\cr}$$
It follows that
$$v_h(3,y,3)\;=\;{\bf if}\; y=3\;{\bf then}\;2\;{\bf else}\;3\,.$$
But we also must have
$$\eqalign{v_h(4,3,1)&=v_h\bigl(v_h(3,3,1),\,v_h(2,1,4),\,v_h(0,4,3)\bigr)\cr
\noalign{\smallskip}
&=v_h\bigl(3,\,v_h\bigl(v_h(1,1,4),\,v_h(0,4,2),\,v_h(3,2,1)\bigr),3\bigr)\cr
\noalign{\smallskip}
&=v_h\bigl(3,\,v_h\bigl(4,3,v_h\bigl(v_h(2,2,1),\,%
v_h(1,1,3),\,v_h(0,3,2)\bigr)\bigr),\,3\bigr)\cr
\noalign{\smallskip}
&=v_h\bigl(3,\,v_h\bigl(4,3,v_h(2,3,3)\bigr),\,3\bigr)\cr
\noalign{\smallskip}
&=v_h\bigl(3,\,v_h(4,3,1),\,3\bigr)\,.\cr}$$
And there is no $y$ such that $y=v_h(3,y,3)$.

\bigskip\noindent
{\bf Open Problem 3.}\enspace If we restrict $h(x,y,z)$ to be strictly
less than $\max(x,y,z)$, is there always a total function $v_h(x,y,z)$
that satisfies the generalized Takeuchi recurrence?\quad\pfbox

\bigskip
We have considered Takeuchi's special case $h(x,y,z)=y$ 
as well as  Gabriel's
special case ${h(x,y,z)=z}$, so it is natural to consider also the
recurrence with $h(x,y,z)=x$. Let
$$\eqalign{k(x,y,z)=\;&{\bf if}\; x\leq y\;{\bf then}\; x\cr
&{\bf else}\; k\bigl(k(x-1,y,z),\,k(y-1,z,x),\,k(z-1,x,y)\bigr)\,.\cr}$$
This recursive definition yields only a partial function because, for
example, we have
$$k(x+1,x,x)=k(x,x-1,x-1)=k(x-1,x-2,x-2)=\cdots\;.$$
However, there are infinitely many ways to define a total function
that does satisfy the $k$~recurrence:

\proclaim Theorem 3. Let $c$ be any integer. The function
$$\eqalign{k_c(x,y,z)=\;&{\bf if}\; x\leq y\;{\bf then}\;x\cr
&{\bf else\ if}\; y\leq z+1\;{\bf then}\; c\;\,{\bf
else}\;\min(y,c)\cr}$$
satisfies the generalized Takeuchi recurrence stated above for
$k(x,y,z)$.

\noindent{\it Proof}.\enspace
Notice that we have the special values
$$\eqalign{&k_c(x,c,z)=\min(x,c)\,;\cr
&{\bf if}\;x>y\;{\bf then}\;k_c(x,y,c)=c\,.\cr}$$
The proof is now by induction on $x-y$.

If $x=y+1$ we have
$$k_c\bigl(k_c(x-1,y,z),\,k_c(y-1,z,x),\,k_c(z-1,x,y)\bigr)
=k_c\bigl(y,\,k_c(y-1,z,y+1),\,k_c(z-1,y+1,y)\bigr)\,.$$
If $y\leq z+1$, this reduces to
$$k_c(y,y-1,z-1\;{\rm or}\; c)=c\,;$$
and if $y\geq z+2$, it is
$$k_c\bigl(y,\,c,\,k_c(z-1,y+1,y)\bigr)=\min(y,c)\,.$$
Thus we obtain $k_c(x,y,z)$ when $x=y+1$.

If $x\geq y+2$ and $y\leq z+1$ we have
$$k_c\bigl(k_c(x-1,y,z),\,k_c(y-1,z,x),\,k_c(z-1,x,y)\bigr)
=k_c(c,\,y-1,\,z-1\;{\rm or}\; c\;{\rm or}\; x)\,,$$
which equals $c$ since $y-1\leq z-1+1$ and $y-1\leq x+1$.

And finally if $x\geq y+2$ and $y\geq z+2$ the right side of the
recurrence reduces to
$$k_c\bigl(\min(y,c),c,z-1\bigr)=\min(y,c)\,.\quad\pfbox$$

\proclaim Corollary. The least fixed point of the recursive definition
$k(x,y,z)$ is {\bf if} $x\leq y$ {\bf then} $x$ {\bf else}~$\omega$.

\noindent{\it Proof}.\enspace
Whenever $x>y$, we have $k_c(x,y,z)=y$ when $c=y$ but not when
$c<y$.\quad\pfbox 

\bigskip\noindent
{\bf 4. The Takeuchi recurrence in higher dimensions.}\enspace
If we define
$$\eqalign{t(w,x,y,z)=\;&{\bf if}\; w\leq x\;{\bf then}\; x\cr
&{\bf else}\;t\bigl(t(w-1,x,y,z),\,t(x-1,y,z,w),\,t(y-1,z,w,x),\,
t(z-1,w,x,y)\bigr)\cr}$$
it turns out that the function reduces to the simple mapping
$$\eqalign{t(w,x,y,z)=\;&{\bf if}\; w\leq x\;{\bf then}\; x\;
{\bf else \ if }\; x\leq y\;{\bf then}\; y\cr
&{\bf else\ if}\;y\leq z\;{\bf then}\; z\;{\bf else}\; w\,.\cr}$$
Therefore it is natural to conjecture that the $m$-dimensional
generalization
$$\eqalign{t(x_1,x_2,\ldots,x_m)=\;&{\bf if}\;x_1\leq x_2\;{\bf
then}\;x_2\cr
&{\bf
else}\;t\bigl(t(x_1-1,x_2,\ldots,x_m),\,\ldots,\,t(x_m-1,x_1,\ldots,
x_{m-1})\bigr)\cr}$$
is satisfied by the $m$-dimensional ``first rise'' function
$$\eqalign{u(x_1,x_2,\ldots,x_m)=\;&{\bf if}\;x_1>\cdots >x_k\leq
x_{k+1}\;{\rm for\ some}\;k\geq 1\;{\bf then}\; x_{k+1}\cr
&{\bf else}\;x_1\,.\cr}$$
But this is false, for all $m>4$. Indeed, we have a 5-dimensional
counterexample,
$$\eqalign{t(5,3,2,0,1)&=t\bigl(t(4,3,2,0,1),\,2,\,t(1,0,1,5,3),\,1,\,
5\bigr)\cr
\noalign{\smallskip}
&=t\bigl(t\bigl(3,2,t(1,0,1,4,3),1,4\bigr),\,2,\,t(0,1,\ldots\,),\,
1,\,5\bigr)\cr
\noalign{\smallskip}
&=t\bigl(t\bigl(3,2,t(0,1,\ldots\,),1,4\bigr),\,2,\,1,\,1,\,5\bigr)\cr
\noalign{\smallskip}
&=t\bigl(t(3,2,1,1,4),\,2,\,1,\,1,\,5\bigr)\cr
\noalign{\smallskip}
&=t\bigl(t(2,1,1,4,3),\,2,\,1,\,1,\,5\bigr)\cr
\noalign{\smallskip}
&=t\bigl(t(1,1,4,3,2),\,2,\,1,\,1,\,5\bigr)=t(1,2,1,1,5)=2\,,\cr}$$
while $u(5,3,2,0,1)=1$.

The true general behavior is somewhat complicated, although
(fortunately) the
complications do not get worse and worse as $m$~grows larger and
larger. Let us define an auxiliary set of functions
$g_j(x_1,\ldots,x_j)$
for $j\geq 2$ as follows:
$$\eqalign{g_j(x_1,\ldots,x_j)=\;&{\bf if}\;j=2\;{\bf then}\;x_2\cr
&{\bf else\ if} \; x_1=x_2+1\;{\bf then}\;g_{j-1}(x_2,\ldots,x_j)\cr
&{\bf else\ if}\;x_2=x_3+1\;{\bf then}\;\max(x_3,x_j)\;{\bf else}\;
x_j\,.\cr}$$

\proclaim Theorem 4. The function
$$\eqalign{f(x_1,\ldots,x_m)=\;&{\bf if}\;x_1>\cdots >x_k\leq x_{k+1}\;
{\rm for\ some}\;k\geq 1\cr
&{\bf then}\;g_{k+1}(x_1,\ldots,x_{k+1})\;{\bf else}\;x_1\cr}$$
satisfies then $m$-dimensional Takeuchi recurrence.

\noindent{\it Proof}.\enspace
Given $x_1,\ldots,x_m$, with $x_1>x_2$, let
$$y_j=f(x_j-1,x_{j+1},\ldots,x_m,x_1,\ldots,x_{j-1})\,.$$
We want to show that $f(y_1,\ldots,y_m)=f(x_1,\ldots,x_m)$.

If $x_1>\cdots >x_m$, we have $y_1=x_1-1$ or $x_2$; $y_2=x_1$ or~$x_3$;
$\ldots\,$; $y_{m-1}=x_1$ or~$x_m$; and $y_m=x_1$. We cannot have
$y_1>\cdots >y_m$, because $y_m>y_1$. Hence there is a unique 
$k\geq 1$ such that $y_1>\cdots >y_k\leq y_{k+1}$. And this can
happen only if $y_{k+1}=x_1$; otherwise $y_{k+1}=x_{k+2}<y_k$.
It follows that $f(y_1,\ldots,y_m)=g_{k+1}(y_1,\ldots,y_{k+1})
=x_1=f(x_1,\ldots,x_m)$.

Assume now that $x_1>\cdots >x_k\leq x_{k+1}$, where $k\geq 2$, and
let $a$ be as large as possible such that $x_i=x_{i+1}+1$ for $1\leq
i<a$. Then $y_i=x_{i+1}$ for $1\leq i<a$. If $a=k$, we have
$y_k=x_{k+1}$ and $f(y_1,\ldots,y_m)=g_k(y_1,\ldots,y_k)=g_k(x_2,\ldots,x_{k+1})=g_{k+1}(x_1,\ldots,x_{k+1})=f(x_1,\ldots,x_m)$.

Assume therefore that $a<k$; hence $x_a>x_{a+1}+1$. Let $b$ be as
large as possible such that $x_i>x_{i+1}+1$ for $a\leq i<b$. 
If $b=k$, we have $y_i=x_{k+1}$ for $a\leq i\leq b$, hence
$f(y_1,\ldots,y_m)=g_{a+1}(y_1,\ldots,y_{a+1})=g_{a+1}(x_2,\ldots,
x_a,x_{k+1},x_{k+1})=x_{k+1}=g_{k+1}(x_1,\ldots,x_{k+1})=
f(x_1,\ldots,x_m)$.

Assume therefore that $b<k$; hence $x_b=x_{b+1}+1$ and $y_b=x_{b+1}$.
Let $z=\max(x_{b+1},x_{k+1})$. We have $y_i=z$ for $a\leq i<b-1$.
If $a>1$ and $z\geq x_a$, we have $f(y_1,\ldots,y_m)=g_a(y_1,\ldots,
y_a)=g_a(x_2,\ldots,x_a,z)=z=g_{k+1}(x_1,\ldots,x_{k+1})=
f(x_1,\ldots,x_m)$.

Assume therefore that $a=1$ or $z<x_a$. If $x_{b-1}>x_b+2$ then
$y_{b-1}=z$; otherwise $y_{b-1}=x_l\geq x_{k+1}$ for some~$l$ in the
range $b+1<l\leq k+1$. If $b>a+2$, we have $f(y_1,\ldots,y_m)=
g_{a+1}(y_1,\ldots,y_{a+1})=g_{a+1}(x_2,\ldots,x_a,z,z)=z=g_{k+1}
(x_1,\ldots,x_{k+1})=f(x_1,\ldots,x_m)$. If $b=a+2$, the same chain
of equalities is valid unless $x_{b-1}=x_b+2$ and $y_{b-1}=x_l$ and
$x_l\not= z$. In the latter case we cannot have $z=x_{k+1}$, for that
would imply $x_{k+1}\geq x_{b+1}\geq x_l$, hence ${x_l=x_{k+1}=z}$.
It follows that $z=x_{b+1}>x_{k+1}$, and $x_l<z=y_b$. Then
$f(y_1,\ldots,y_m)=g_{a+2}(y_1,\ldots,y_{a+2})=g_{a+2}(x_2,\ldots,
x_a,z,x_l,z)=z=f(x_1,\ldots,x_m)$.

Assume therefore that $b=a+1$. If $z=x_{k+1}$ we have $x_{k+1}\geq
x_i$ for $b<i\leq k$, hence $y_a=z$ and $y_i=x_{i+1}$ or~$z$ for
$a<i\leq k$. The first appearance of~$z$ among $y_{a+1},\ldots,
y_{k+1}$ will show that $f(y_1,\ldots,y_m)=z=f(x_1,\ldots,x_m)$.

Assume therefore that $z=x_{b+1}>x_{k+1}$. If $x_a>x_b+2$ then
$y_a=y_b=z$, hence $f(y_1,\ldots,y_m)=g_b(x_2,\ldots,x_a,z,z)
=z=f(x_1,\ldots,x_m)$.

Assume therefore (and finally) that $x_a=x_b+2$, so that
$y_a=x_l\geq x_{k+1}$, where $b+1<l\leq k+1$. Then $x_l<x_{b+1}
=y_b$, and $f(y_1,\ldots,y_m)=g_b(x_2,\ldots,x_a,x_l,x_{b+1})
=z=f(x_1,\ldots,x_m)$. We have proved that $f(y_1,\ldots,y_m)
=f(x_1,\ldots,x_m)$ in all cases. \quad\pfbox

\bigskip
A machine-based proof of Theorem 4 would be very interesting,
especially
if it could cope with functions having a variable number of arguments.

Notice that we have not proved that the $m$-dimensional  Takeuchi
recursion $t(x_1,\ldots,x_m)$ actually defines a total
function,
when $m>3$. We have only shown that $f(x_1,\ldots,x_m)$ satisfies
the recurrence. If the repeated expansion of $t(x_1,\ldots,x_m)$
actually terminates for some sequence of arguments $(x_1,\ldots,x_m)$,
it must yield the value $f(x_1,\ldots,x_m)$; but we have not 
demonstrated that termination will occur, and there is apparently
no obvious ordering on the integer $m$-tuples $(x_1,\ldots,x_m)$
that will yield such a proof. Therefore we come to a final question,
which will perhaps prove to be the most interesting aspect of the
present investigation.

\bigskip\noindent
{\bf Open Problem 4.}\enspace
Does the $m$-dimensional Takeuchi recursion equation define a total
function, for all $m\geq 3$, if it is expanded fully (without 
call-by-need)? Equivalently, does the recurrence
$$\eqalign{T(x_1,\ldots,x_m)=\;&{\bf if}\;x_1\leq x_2\;{\bf then}\;0\cr
&{\bf else}\;1+T(x_1-1,x_2,\ldots,x_m)+T(x_2-1,x_3,\ldots,x_m,x_1)+
\cdots\cr
&\phantom{{\bf else}\;1}
\null+T(x_{m-1}-1,x_m,x_1,\ldots,x_{m-2})+T(x_m-1,x_1,\ldots,x_{m-1})\cr
&\phantom{{\bf else}\;1}
\null+T\bigl(f(x_1-1,x_2,\ldots,x_m),\ldots,f(x_m-1,x_1,\ldots,x_{m-1})
\bigr)\cr}$$
define a total function on the integers $(x_1,\ldots,x_m)$, for all
$m\geq 3$? (Here $f$~is the function of Theorem~4.)\quad\pfbox

\bigskip
Close inspection of the proof of Theorem 4 implies that a 
call-by-need technique
{\it will\/} always terminate when applied to the
recursive equation for $t(x_1,\ldots,x_m)$. If $x_1>x_2>\cdots 
>x_k\leq x_{k+1}$, the values $y_i=t(x_i-1,x_{i+1},\ldots,x_{i-1})$
need be expanded only for $1\leq i\leq k+1$, and this will be sufficient
to determine the value of $t(y_1,\ldots,y_m)=t(x_1,\ldots,x_m)$ in a
finite number of steps. (The proof is by induction on~$k$.) However,
the possibility remains that an attempt to expand the ``irrelevant''
parameters $y_{k+2},\ldots,y_m$ might loop forever. If so, the
Takeuchi recurrence would be an extremely interesting example to
include in {\it all\/} textbooks about recursion.

\bigskip
\centerline{\bf References}

\parskip 3pt

\medskip

\bib
\phantom{1}[1]\enspace
N. G. de Bruijn, {\sl Asymptotic Methods in Analysis\/} 
(Amsterdam: North-Holland, 1961).

\bib
\phantom{1}[2]\enspace
Richard P. Gabriel, {\sl Performance and Evaluation of Lisp Systems\/}
(Cambridge, Mass.: MIT Press, 1985).

\bib
\phantom{1}[3]\enspace
Ronald L. Graham, Donald E. Knuth, and Oren Patashnik, {\sl Concrete
Mathematics\/} (Reading, Mass.: Addison\kern.1em--Wesley, 1989).

\bib
\phantom{1}[4]\enspace
Donald E. Knuth, {\sl Fundamental Algorithms\/} (Reading, Mass.:
Addison\kern.1em--Wesley, 1968).

\bib
\phantom{1}[5]\enspace
Donald E. Knuth and Herbert S. Wilf, ``A short proof of Darboux's
lemma,'' {\sl Applied Mathematics Letters\/ \bf 2} (1989), 139--140.

\bib
\phantom{1}[6]\enspace
Zohar Manna, {\sl Mathematical Theory of Computation\/} (New York:
McGraw-Hill, 1974).

\bib
\phantom{1}[7]\enspace
Zohar Manna and John McCarthy, ``Properties of programs and partial
function logic,'' {\sl Machine Intelligence\/ \bf 5} (1970), 27--37.

\bib
\phantom{1}[8]\enspace
Zohar Manna and Amir Pnueli, ``The validity problem of the
91-function,''
Stanford Artificial Intelligence Project, Memo No.~68 (August 19,
1968), 20~pp.

\bib
\phantom{1}[9]\enspace
Zohar Manna and Amir Pneuli, ``Formalization of properties of
functional programs,'' {\sl Journal of the ACM\/ \bf 17} (1970), 555--569.

\bib
[10]\enspace
John McCarthy, ``An interesting LISP function,'' unpublished
notes, autumn 1978, 3~pp.

\bib
[11]\enspace
John McCarthy and Carolyn Talcott, {\sl Lisp: Programming and Proving}.
Course notes, Computer Science Department, Stanford University, 1980.
``Under revision for publication as a book.''

\bib
[12]\enspace
J Strother Moore, ``A mechanical proof of the termination of
Takeuchi's function,'' {\sl Information Processing Letters\/ \bf 9}
(1979), 176--181.

\bib
[13]\enspace
I. Takeuchi, ``On a recursive function that does almost recusion
only,'' Electrical Communication Laboratory, Nippon Telephone and 
Telegraph Co., Tokyo, Japan (1978).

\bib 
[14]\enspace
Ilan Vardi, ``Running time of TAK,'' preliminary version of
unpublished manuscript dated December~1988.

\bib
[15]\enspace
Jean Etienne Vuillemin, Proof techniques for recursive programs, Ph.D.
thesis, Stanford University, 1973.

\bib
[16]\enspace
Jean Vuilllemin, ``Correct and optimal implementations of recursion in
a simple programming language,'' {\sl Fifth Annual ACM Symposium on Theory
of Computing\/} (1973), 224--239.

\bigskip\bigskip

The preparation of this paper was supported in part by National
Science Foundation grant CCR--8610181.

\bye